\documentclass[article, twocolumn,
  amsmath,amssymb,
  longbibliography,
  ]{revtex4-1}
\usepackage{graphicx,color}
\usepackage{amsmath}
\usepackage{natbib}
\usepackage{epsfig}
\begin{document}

\title{On the system size dependence of the diffusion coefficients in MD simulations: A simple correction formula for pure dense fluids}

\author{S. A. Khrapak}\email{Sergey.Khrapak@gmx.de}
\affiliation{Joint Institute for High Temperatures, Russian Academy of Sciences, 125412 Moscow, Russia}

\begin{abstract}
A practical correction formula relating the self-diffusion coefficient of dense liquids from molecular dynamics simulations with periodic boundary conditions to the self-diffusion coefficient in the thermodynamic limit is discussed. This formula applies to pure dense fluids and has a very simple form $D=D_0(1-\gamma N^{-1/3})$, where $D_0$ is the self-diffusion coefficient in the thermodynamic limit and $N$ is the number of particles in the simulation. The numerical factor $\gamma$ depends on the geometry of the simulation cell. Remarkably, $\gamma\simeq 1.0$ for the most popular cubic geometry. The success of this formula is supported by results from MD simulations, including very recent simulations with a ``magic'' simulation geometry.         
\end{abstract}

\date{\today}

\maketitle

\section{Introduction}

It is known that the self-diffusion coefficient obtained using molecular dynamics (MD) simulations with a finite number of particles may exhibit considerable system size dependence. This effect has been successfully interpreted in terms of hydrodynamic interactions between the particles. For a cubic box of size $L$ with periodic boundary conditions a correction formula
\begin{equation}\label{corr_cubic}
D=D_0-\frac{k_{\rm B}T\xi}{6\pi \eta L}
\end{equation}
was derived~\cite{DunwegJCP1993,YehJPCB2004}. Here $D$ is the diffusion coefficient obtained in a simulation, $D_0$ is the diffusion coefficient in the thermodynamic limit, $k_{\rm B}$ is the Boltzmann constant, $T$ is the temperature, $\eta$ is the shear viscosity coefficient, and $\xi$ is a numerical factor, which represents the analogue of a Madelung constant in the Ewald summation~\cite{DunwegJCP1993}.  

Kikugawa with co-workers generalized this consideration to rectangular geometry. They first showed that in the rectangular
box system, the diffusion tensor appears anisotropic even for isotropic fluids~\cite{KikugawaJCP2015}. In a subsequent paper~\cite{KikugawaJCP2015_1} the  hydrodynamic theory applied to periodic rectangular box systems yielded a correction of the form
\begin{equation}\label{corr_rectangular}
D_{jj}= D_0-\frac{k_{\rm B}T\xi_{jj}}{6\pi \eta L_j},
\end{equation}
where $j = (x,y,z)$, $D_{jj}$ are the diagonal terms of the diffusion tensor, and $L_j$ is the length of the box in the corresponding direction. The parameters $\xi_{jj}$ can be expressed in the form of a summation over real and reciprocal lattice vectors~\cite{KikugawaJCP2015_1}.   

The purpose of this paper is to report a simplification of the above formulas, which allows to express the correction to the diffusion coefficient in terms of only the number of particles employed in the numerical simulation. 

\section{Methods}

The correction formulas (\ref{corr_cubic}) and (\ref{corr_rectangular}) can be greatly simplified in a dense fluid regime, where Stokes-Einstein (SE) relation without the hydrodynamic diameter holds. The SE relation can be written as 
\begin{equation}\label{SErel}
D_0\eta\left(\frac{\Delta}{k_{\rm B}T}\right)=\alpha_{\rm SE},
\end{equation}
where $\Delta=\rho^{-1/3}$ is the mean inter-atomic (or inter-molecular) separation and $\alpha_{\rm SE}$ is the SE coefficient, which may depend on the fluid properties as well as on the exact location on the phase diagram. However, this dependence is rather weak, as discussed below.

Originally, SE relation without the hydrodynamic diameter was mainly discussed in the context of simple fluids~\cite{FrenkelBook,BalucaniBook,ZwanzigJCP1983,
Balucani1990,CostigliolaJCP2019}. Zwanzig~\cite{ZwanzigJCP1983} provided a particularly appealing theoretical demonstration regarding why a relation of the kind of Eq.~(\ref{SErel}) should be expected to work in simple fluids. He also derived the lower and upper bounds on the SE coefficients, $0.132\lesssim \alpha_{\rm SE}\lesssim 0.181$. 

Numerous confirmations of the applicability of the microscopic SE relation in the form of Eq.~(\ref{SErel}) to dense simple model fluids have been reported~\cite{KhrapakPR2024}. These include single component Coulomb (one-component plasma) and screened Coulomb (complex or dusty plasma) fluids of charged particles~\cite{DaligaultPRL2006,DaligaultPRE2014,KhrapakAIPAdv2018,
KhrapakMolecules12_2021,KhrapakPRE10_2021}, soft (inverse power) repulsive particle fluid~\cite{HeyesPCCP2007}, Lennard-Jones fluid~\cite{CostigliolaJCP2019,OhtoriPRE2015,OhtoriPRE2017,
KhrapakPRE10_2021}, Weeks-Chandler-Andersen fluid~\cite{OhtoriJCP2018}, and the hard sphere fluid~\cite{OhtoriJCP2018,Pieprzyk2019,KhrapakPRE10_2021}.
As it becomes more and more evident, the applicability of SE relation is not limited to simplest point-like and monatomic models with isotropic interactions. 
Several important non-spherical molecular liquids have been examined using numerical simulations in Ref.~\cite{OhtoriChemLett2020} and the applicability  of the SE relation has been confirmed. 
A few recent examples confirming the applicability of SE relation to real liquids include liquid iron at
conditions of planetary cores~\cite{LiJCP2021}, dense supercritical methane (at least for the most state points investigated)~\cite{Ranieri2021,KhrapakJMolLiq2022}, silicon melt at high temperatures~\cite{Luo2022}, and liquid water modelled by the TIP4P/Ice model~\cite{BaranJCP2023,KhrapakJCP2023} (which was specifically designed to deal with water near the fluid-solid phase transition and solid-phase properties~\cite{AbascalJCP2005}). 

A typical ''average'' value of the SE coefficient emerging in these studies is $\alpha_{\rm SE}\simeq 0.15$. It is systematically lower ($\alpha_{\rm SE}\simeq 0.14$) for plasma-related systems with extremely soft interparticle interactions (e.g. of Coulomb type) and systematically higher ($\alpha_{\rm SE}\simeq 0.17$) for extremely steep hard-sphere interaction~\cite{KhrapakPRE10_2021}. This can be rationalized in terms of the effect of interaction potential steepness on the fluid instantaneous elastic moduli as explained in Ref.~\cite{KhrapakMolPhys2019}. However, if we just adopt the characteristic value of $\simeq 0.15$ for $\alpha_{\rm SE}$, then we can rewrite Eq.~(\ref{corr_cubic}) in a particularly simple form as  
\begin{equation}\label{corr_cubic_N}
D\simeq D_0(1-N^{-1/3}),
\end{equation}                     
where the identity $\rho L^3=(L/\Delta)^3=N$ has been used and the numerical factor appropriate for a cubic simulation box~\cite{DunwegJCP1993}, $\xi\simeq 2.8373$, has been taken. This coincides to within the first order correction terms with the expression proposed in Ref.~\cite{KhrapakMolPhys2019}. Yeh and Hummer also used the SE relation for an approximate estimate of the systems size correction~\cite{YehJPCB2004}. However, they did not use the form of the SE relation without the hydrodynamic radius. Instead, their approach required an estimate of the hydrodynamic radius as well as the choice of stick or slip boundary condition. 

Equation (\ref{corr_cubic_N}) can be easily generalized to rectangular geometry. Assume for instance the simulation box with $L_x=L_y\neq L_z$. Then the diagonal $z$-component of the diffusion tensor becomes
\begin{equation}\label{corr_rectangular_N}
D_{zz}=D_0(1-\gamma N^{-1/3}),
\end{equation}               
where a geometry-dependent coefficient $\gamma$ is expressed as $\gamma = (\xi_{zz}/6\pi \alpha_{\rm SE})(L_{x}/L_{z})^{2/3}$. For a known box length ratio the coefficient $\gamma$ can be calculated analytically. 

\section{Results and discussion}

In a recent paper Busch and Paschek elaborated on an interesting proposal to use a rectangular system with a ``magic'' box length ratio $L_{\rm z}/L_{x}=L_z/L_y\simeq 2.7934$ to compute simultaneously the self-diffusion and shear viscosity coefficients~\cite{BuschJPCB2023}. For this specific geometry, the self-diffusion coefficients in the $x$ and $y$ directions become system-size independent (simply because $\xi_{xx}=\xi_{yy}\simeq 0$ to a good accuracy). Therefore the ``true'' self-diffusion coefficient in the thermodynamic limit can be simply defined as $D_0=(D_{xx}+D_{yy})/2$. Measuring the $z$-component of the diffusion tensor one can then evaluate the shear viscosity coefficient from Eq.~(\ref{corr_rectangular}). A relative accuracy in estimating $\eta$ (of about $\sim 10\%$) is not particularly impressive, but is nevertheless comparable with other methods such as integration of the stress-tensor autocorrelation function. 

\begin{figure}
\includegraphics[width=7.8cm]{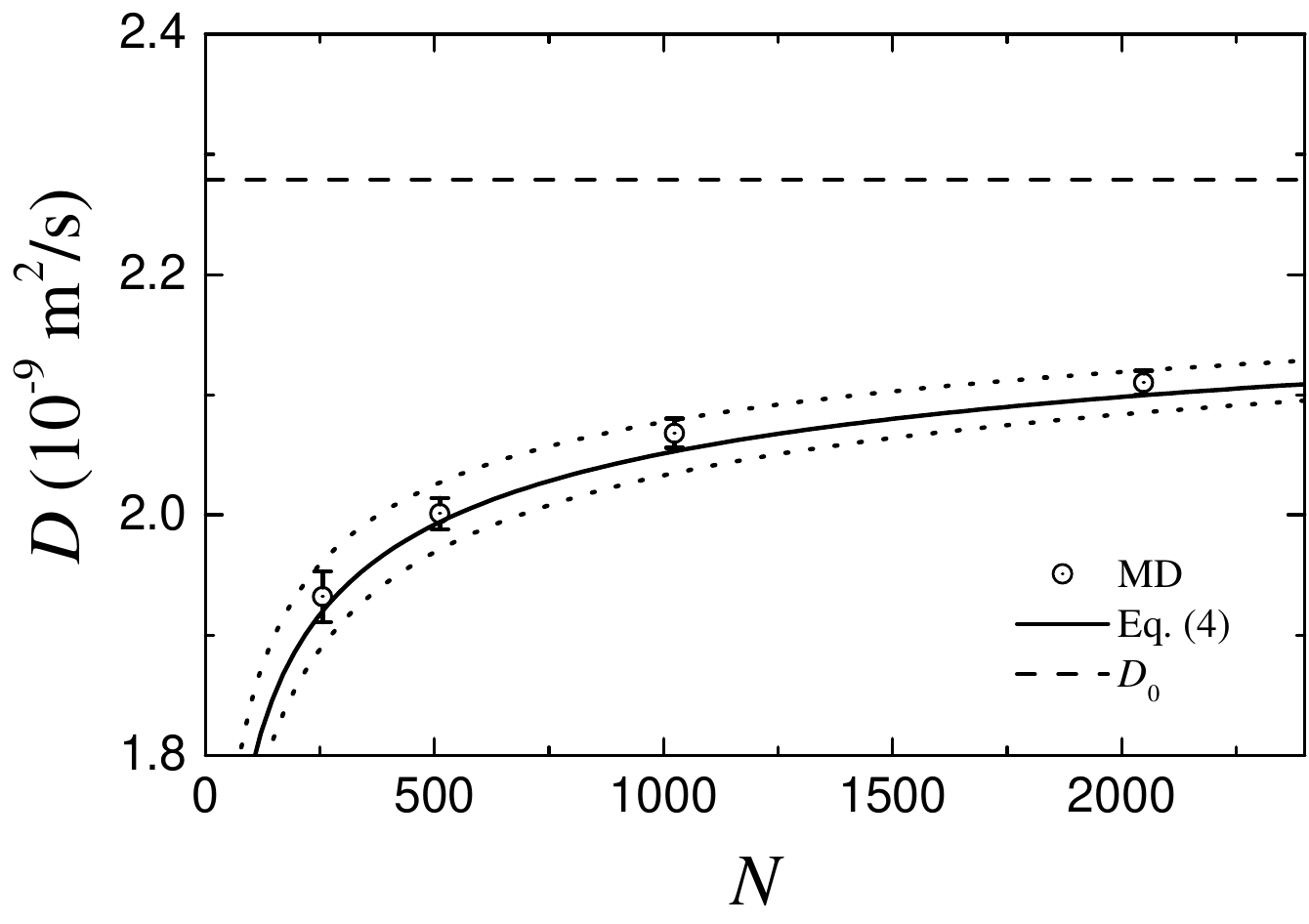}
\caption{Self-diffusion coefficient of TIP4P/2005 water model at $T= 298$ K vs the number of particles $N$ for an MD simulation with a cubic box, $L_x=L_y=L_z$. The symbols denote MD results for a finite number of particles (Table II from Ref.~\cite{BuschJPCB2023}). The dashed curve marks the computed thermodynamic limit value $D_0\simeq 2.28\times 10^{-9}$ m$^2$/s. The solid curve corresponds to Eq.~(\ref{corr_cubic_N}). The doted curves correspond to the same functional form, but with $\alpha_{\rm SE}=0.14$ and $0.17$.}
\label{Fig1}
\end{figure} 

To support their approach (referred to as OrthoBoXY), Busch and Paschek performed molecular dynamics simulations using the TIP4P/2005 model of liquid water~\citep{AbascalJCP2005/TIP4P2005}. Simulations were performed for $NVT$ and $NPT$ ensembles at $T= 298$ K
and $\rho_{\rm m}=0.9972$ g/cm$^{3}$ ($NVT$) and $P=1$ bar ($NPT$). Various system sizes were used for cubic and rectangular geometries. Further details regarding MD simulation protocol can be found in Busch and Paschek paper~\cite{BuschJPCB2023}. They demonstrated the feasibility of their method to determine simultaneously the diffusion and viscosity coefficients from a single MD simulation run. The obtained self-diffusion and shear viscosity coefficients agree reasonably well with the experimental data. 

\begin{figure}
\includegraphics[width=7.8cm]{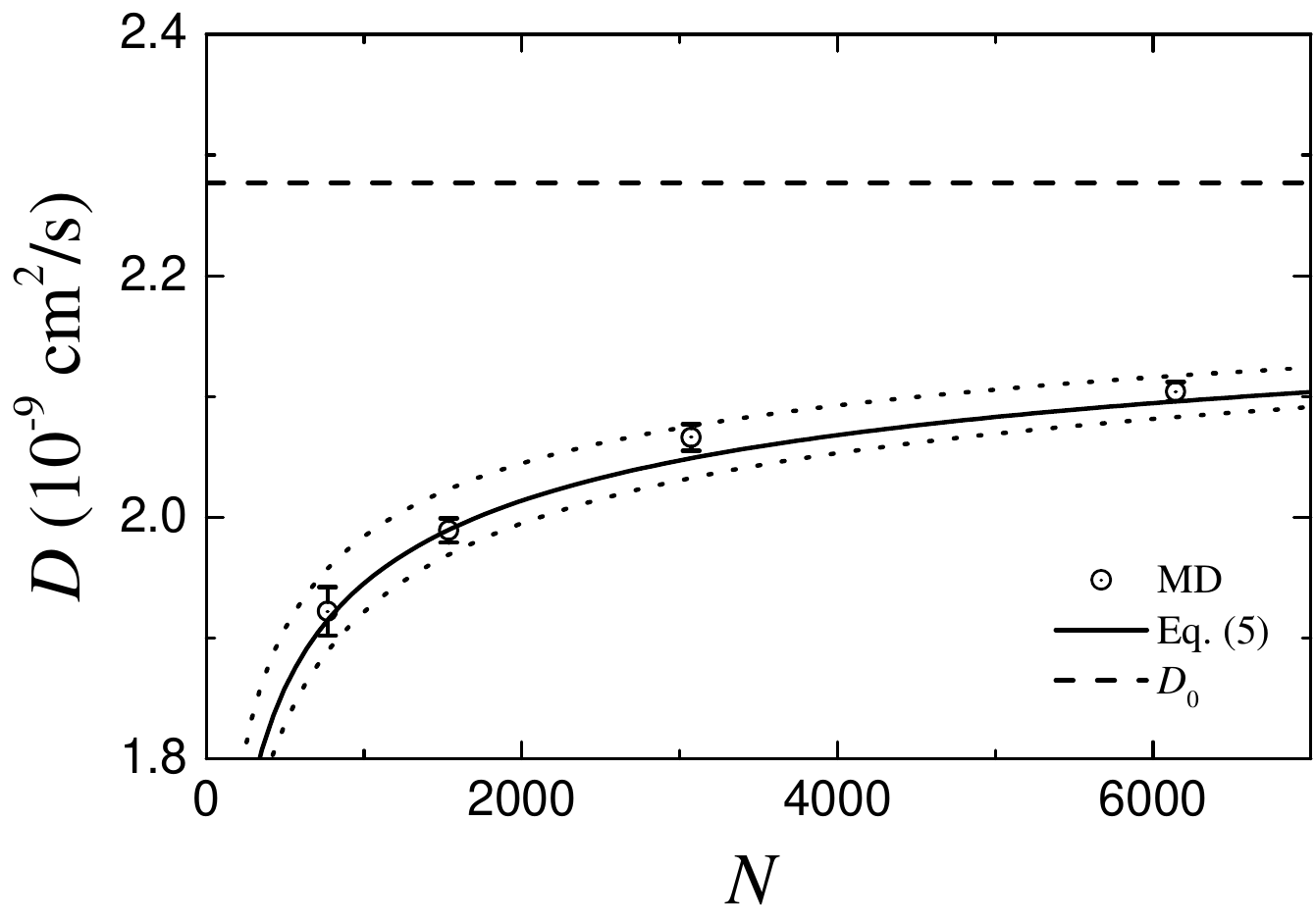}
\caption{Self-diffusion coefficient of TIP4P/2005 water model at $T=  298$ K vs the number of particles $N$ for an MD simulation with a ``magic'' box-length ratio $L_z/L_x=L_z/L_y\simeq 2.7934$.  The symbols denote the $D_{zz}$ element of the diffusion tensor (Table I from Ref.~\cite{BuschJPCB2023}). The dashed curve marks the computed thermodynamic limit value $D_0\simeq 2.28\times 10^{-9}$ m$^2$/s. The solid curve corresponds to Eq.~(\ref{corr_rectangular_N}) with $\gamma\simeq 1.457$. The doted curves correspond to the same functional form, but with $\alpha_{\rm SE}=0.14$ and $0.17$.}
\label{Fig2}
\end{figure} 

Looking from a somewhat different perspective, the results by Busch and Paschek provide an excellent opportunity to verify the accuracy of simple practical correction formulas (\ref{corr_cubic_N}) and (\ref{corr_rectangular_N}). 

Figure~\ref{Fig1} shows comparison between MD results and Eq.~(\ref{corr_cubic_N}) for simulations performed in a cubic box under $NVT$ conditions (see Table II from Ref.~\cite{BuschJPCB2023}). Very good agreement is observed. In order to give an idea how robust is the choice of $\alpha_{\rm SE}=0.15$, leading to Eq.~(\ref{corr_cubic_N}), for different systems I also plot the curves corresponding to the soft and hard interaction limits, $\alpha_{\rm SE}=0.14$ and $\alpha_{\rm SE}=0.17$ (dotted curves in Fig.~\ref{Fig1}). Although the correction formula is not very sensitive to the exact choice of $\alpha_{\rm SE}$, knowledge of the SE coefficients for a particular fluid under consideration would be certainly beneficial. Fig.~\ref{Fig2} provides a comparison between MD results and Eq.~(\ref{corr_rectangular_N}) for simulations performed in a ``magic'' box with $L_z/L_x=L_z/L_y\simeq 2.7934$. For this geometry the directional coefficient $\xi$ is $\xi_{zz}\simeq 8.1711$, which yields $\gamma\simeq 1.457$ in Eq.~(\ref{corr_rectangular_N}). An excellent agreement between simple correction formula and MD simulation is again documented. As in Fig.~\ref{Fig1}, the dotted curves correspond to the soft and hard interaction limits, $\alpha_{\rm SE}=0.14$ and $\alpha_{\rm SE}=0.17$.    

{\it A posteriori}, taking simulation values $D_{0}\simeq 2.28\times 10^{-9}$ m$^2$/s and $\eta\simeq 0.90$ mPa s along with $\rho_{\rm m}\simeq 0.997$ g/cm$^3$ and $T= 298$K we obtain $\alpha_{\rm SE}\simeq 0.155$, very close to an ``average'' value adopted above.  

\begin{figure}
\includegraphics[width=7.8cm]{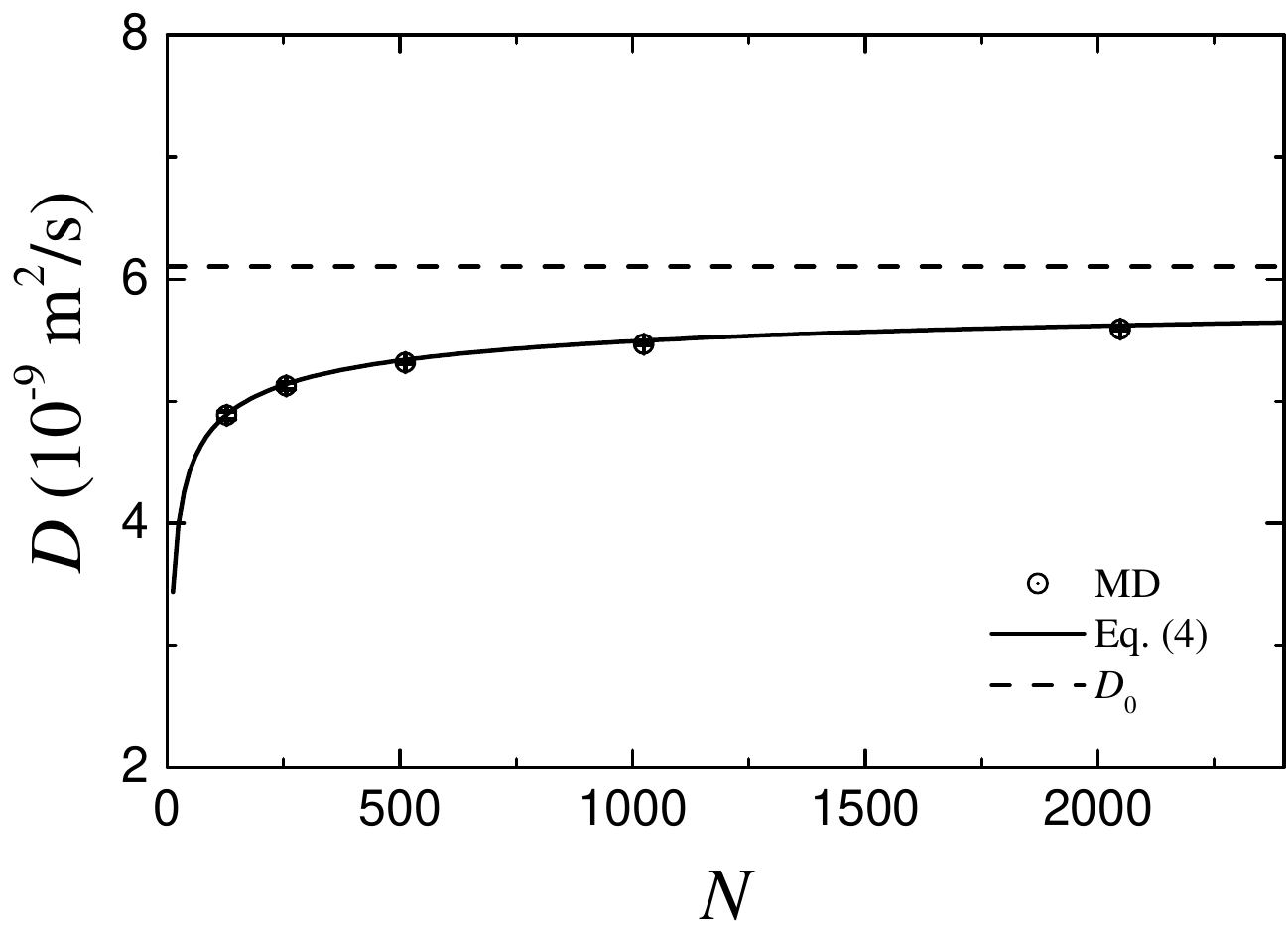}
\caption{System-size dependence of the self-diffusion coefficient in TIP3P water at a temperature $298$ K and near-ambient pressure. The symbols correspond to numerical results from simulations in a cubic simulation box (see Table 1 from Ref.~\cite{YehJPCB2004}). The dashed horizontal line denotes the diffusion coefficient in the thermodynamic limit, $D_0\simeq 6.1\times 10^{-9}$ m$^2$/s. The solid curve corresponds to Eq.~(\ref{corr_cubic_N}).   }
\label{Fig4}
\end{figure} 

\begin{figure}
\includegraphics[width=7.8cm]{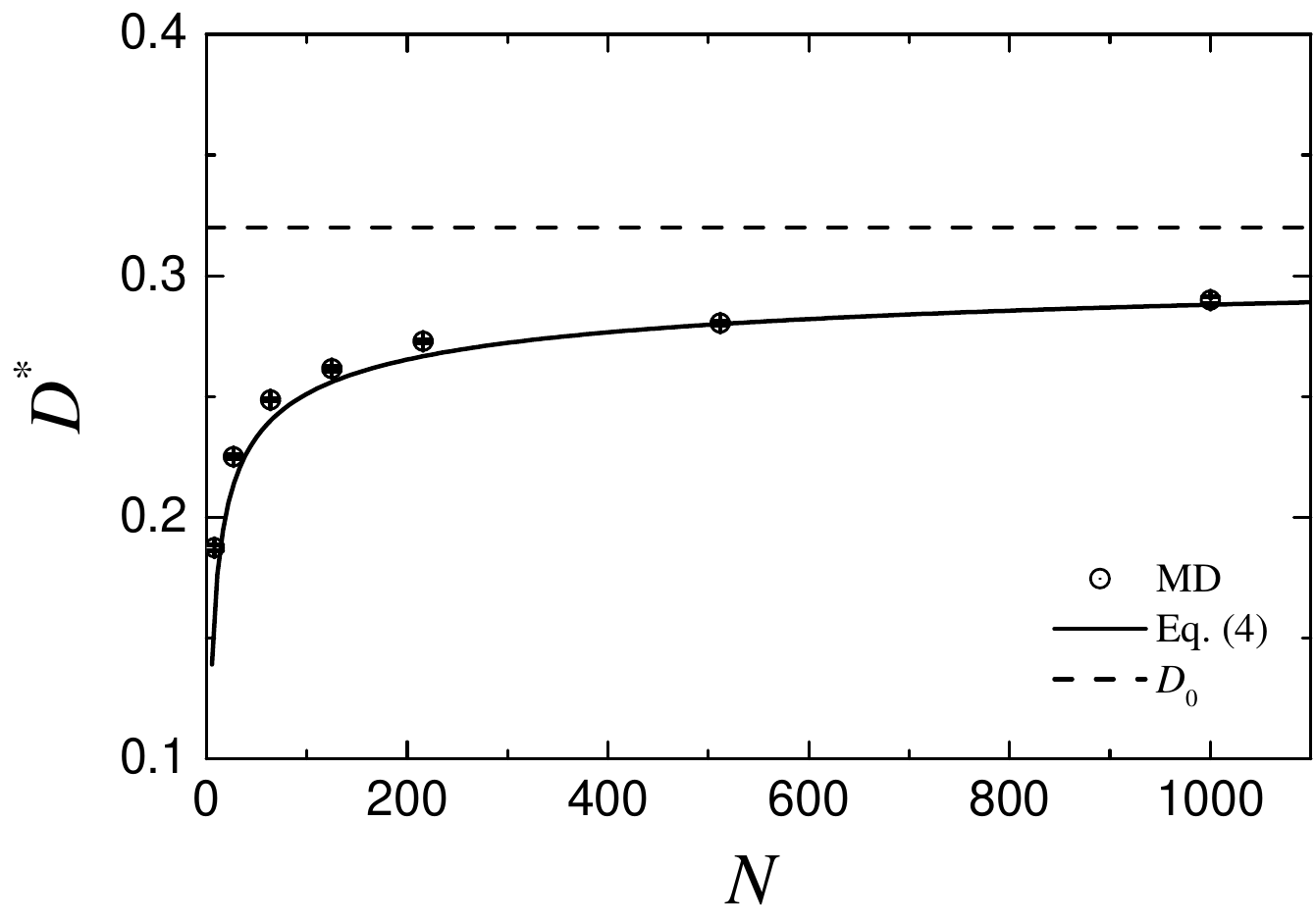}
\caption{System-size dependence of the self-diffusion coefficient of the Lennard-Jones (LJ) fluid at a reduced temperature $T^*=2.75$ and a reduced density $\rho^*=0.7$. The symbols correspond to numerical results from simulations in a cubic simulation box (see Table 2 from Ref.~\cite{YehJPCB2004}). The dashed horizontal line denotes the diffusion coefficient in the thermodynamic limit, $D^*_0\simeq 0.32$. The solid curve corresponds to Eq.~(\ref{corr_cubic_N}).   }
\label{Fig3}
\end{figure}

To avoid an impression that we have been dealing with just a fortunate coincidence, let me present another couple of validations. These are based on the simulations reported by Yeh and Hummer in Ref.~\cite{YehJPCB2004}. The simulations were performed in a cubic simulation cell with different number of simulated particles. The first example corresponds to TIP3P model of water~\cite{JorgensenJCP1983} at a temperature $298$ K and density $33.00$ nm$^{-3}$ under near-ambient pressure. In this case the number of simulated molecules varied from $N=128$ to $N=2048$. The results for the self-diffusion coefficient, tabulated in Tab.~1 of Ref.~\cite{YehJPCB2004}, are shown in Fig.~\ref{Fig4} by symbols. The horizontal dashed line is the diffusion coefficient in the thermodynamic limit. The system-size dependence observed in simulations is very well described by Eq.~(\ref{corr_cubic_N}) shown as the solid curve. Note almost three times difference between the self-diffusion coefficients of TIP4P/2005 (Fig.~\ref{Fig2}) and TIP3P (Fig.~\ref{Fig3}) water models. This difference is compensated by about the same ratio of viscosities, so that the SE relation holds in both cases and Eq.~(\ref{corr_cubic_N}) is equally applicable. The second example corresponds to the Lennard-Jones (LJ) fluid at a reduces temperature $T^* = 2.75$ and reduced density $\rho^*=0.7$ (conventional LJ units are used for normalization). This state point belongs to the region of applicability of SE relation, although lies close to its onset (see Fig. 2 from Ref.~\cite{KhrapakPRE10_2021}). Simulations were performed with the number of particles ranging from $N=8$ to $N=1000$~\cite{YehJPCB2004}. Figure~\ref{Fig3} shows the resulting system-size dependence of the self-diffusion coefficient. Symbols correspond to the data tabulated in Tab.~2 of Ref.~\cite{YehJPCB2004}. The dashed line marks the corrected diffusion coefficient in the thermodynamic limit. The solid curve is the theoretical prediction of Eq.~(\ref{corr_cubic_N}). The agreement is again quite convincing, justifying the proposed practical expression.    
 
\section{Conclusion}
    
In this paper, using recently published and earlier data on the self-diffusion coefficient of different fluids obtained from MD simulations with periodic boundary conditions in cubic and rectangular geometry, it has been demonstrated that simple practical expressions of Eqs.~(\ref{corr_cubic_N}) and (\ref{corr_rectangular_N}) can be quite helpful to correct for the effect of finite particle number. These practical expressions can be applied to new simulation data as well as to the results already published in the literature. The obtained results cannot generally replace the original equations by D\"unweg and Kramer and Yeh and Hummer, but can considerably simplify calculations in cases when SE relation without the hydrodynamic diameter is known to hold.    






\bibliography{SE_Ref}

\end{document}